# How to impose stick boundary conditions in coarse-grained hydrodynamics of Brownian colloids and semi-flexible fiber rheology

Robert D. Groot

*Unilever Research Vlaardingen, PO box 114, 3130 AC Vlaardingen, The Netherlands*

Long-range hydrodynamics between colloidal particles or fibers is modelled by the Fluid Particle Model. Two methods are considered to impose the fluid boundary conditions at colloidal surfaces. In the first method radial and transverse friction forces between particle and solvent are applied such that the correct friction and torque follows for moving or rotating particles. The force coefficients are calculated analytically and checked by numerical simulation. In the second method a collision rule is used between colloidal particle and solvent particle that imposes the stick boundary conditions exactly. The collision rule comprises a generalisation of the Lowe-Anderson thermostat to radial and transverse velocity differences.

## I. INTRODUCTION

Simulation of the rheology of colloids and semi-flexible fibers faces two difficulties. At high volume fraction physics is dominated by hydrodynamic lubrication forces and by associative interactions at particle contact. At low volume fractions particles mainly interact via long-range hydrodynamic interactions. Simulation of dilute suspension rheology thus requires a correct representation of these long-range interaction forces.[1] One way of doing this is to solve the Navier-Stokes equation with moving particles as boundary condition for the flow, e.g. by using the Lattice Boltzmann approach[2] or by using (accelerated) Stokesian Dynamics.[3] Another approach is to model the fluid phase by soft repulsive particles with pair-wise noise and friction, known as dissipative particle dynamics (DPD). This method by construction produces full inertial hydrodynamics,[4] but applying the correct fluid-particle boundary condition is non-trivial. Application to suspension rheology was pioneered by Boek *et al.*[5] Their method consisted of binding a number of DPD particles together and submerging these into a fluid of similar, unbound particles. The comparison with experiments was however qualitative.

Martys[6] later used some 2100 (soft) DPD particles bound together to form a colloidal particle; and other DPD particles were used as solvent. These simulations requires very large numbers of particles (up to 2.2 million) just to reproduce the hydrodynamic interactions. To solve this problem, Pryamitsyn and Ganesan[7] explored a different route to simulate hydrodynamics between particles, which requires only a fraction of the number of particles. They used the Fluid Particle Model (FPM), which includes not only radial friction but also shear friction. Qualitatively, this reproduces the increase of viscosity with volume fraction quite well, and it predicts shear-thinning.

However, in this method the effective volume fraction of particles does not match the real volume fraction. This problem can be traced back to incorrect (stick) boundary conditions between fluid and particles. To solve this problem, Pan *et al.*[8] used radial and shear friction forces between colloid and fluid that differed from the values used between fluid particles mutually. However, for a realistic simulation of suspension rheology, this method requires very high values of the friction coefficient, and consequently only small time steps can to be taken.[9] Moreover, the simulation parameters need to be determined numerically.

The preferred method would be to apply the solvent boundary condition directly to the surface without the need to do parameterisation work at forehand. For this purpose we need a collision rule between solvent and colloidal particles that automatically generates stick boundary conditions. Such a collision rule is developed here. This method allows much larger time steps than FPM, and the boundary condition is imposed exactly without the need of a numerical parameterisation. This makes the present method an estimated 100 times faster than FPM, and some five orders of magnitude faster than accelerated Stokesian Dynamics.

The brief outline of this work is as follows. In section II some basic theory of suspension rheology is reviewed. In section III the Fluid Particle Model is briefly described. In section IV two methods are discussed to define the interaction between colloidal particles and the surrounding fluid via their surface forces and stick boundary conditions, and in section V this is applied to Brownian colloids and fiber suspensions. Finally, section VI gives a summary and conclusions.

## II. THEORY OF SUSPENSION RHEOLOGY

First a brief overview of the established theory of suspension rheology is presented. In general, when particles at low dilution are added to a solvent, the viscosity of the suspension is higher than the viscosity of the surrounding solvent. The viscosity increase itself is again proportional to solvent viscosity. Hence at low dilution (see e.g. Bicerano *et al*[10] for a review) we have

$$\eta = \eta_0(1 + [\eta]\phi + k_H\phi^2 + ...) \qquad (1)$$

where $\eta_0$ is the solvent viscosity, $\phi$ is the particle volume fraction and the coefficients of $\phi$ and $\phi^2$ are known as the intrinsic viscosity $[\eta]$ (first calculated for spheres by Einstein[11]) and the Huggins coefficient $k_H$. The Huggins coefficient is quite sensitive to particle-particle interaction, but if we restrict ourselves to particles without mutual attraction all coefficients can be calculated using Krieger-Dougherty theory.[12]

In brief, the Krieger-Dougherty theory starts with a solvent without particles. We add a small volume fraction of particles, which gives a viscosity increase $d\eta = \eta_0[\eta]d\phi$. If we continue to add particles, we have to replace $\eta_0$ by $\eta(\phi)$, the viscosity at finite volume fraction $\phi$, assuming that a fluid with particles can be treated as if it was a homogeneous fluid. If the particles have a hard core interaction, the volume fraction is limited to a maximum value $\phi_{max}$, therefore the available volume for new particles is $V(1-\phi/\phi_{max})$. The volume fraction of newly added particles must be related to this free volume, therefore the viscosity increases as $d\eta = \eta(\phi) [\eta] d\phi/(1-\phi/\phi_{max})$. This equation can be integrated to yield

$$\eta = \eta_0 (1-\phi/\phi_{max})^{-[\eta]\phi_{max}} \quad (2)$$

i.e. we find a viscosity that diverges at the maximum volume fraction. By construction the limiting slope at low volume fraction is the correct intrinsic viscosity, but the equation also predicts the Huggins coefficient as $k_H = \frac{1}{2}[\eta]([\eta]+1/\phi_{max})$ and the coefficients of all other powers of $\phi$.

For spherical solid particles with stick boundary condition the intrinsic viscosity is[11] $[\eta] = 2.5$. For ellipsoidal particles and other shapes the intrinsic viscosity is larger. For many particle shapes analytical results are known.[13] For instance, for dumbbells of two touching spheres the intrinsic viscosity is[13] $[\eta] = 3.4496$. A special class of particles is ellipsoids of revolution, which have axis of symmetry $a$ and two other axes $b=c$ of equal length. For these ellipsoids the aspect ratio $A = a/b$ plays a central role. For prolate ellipsoids (needle shapes, $A \to \infty$), Onsager[14] calculated $[\eta] \sim \frac{4}{15}A^2/\ln(A)$. Bicerano et al.[10] give a list for a wide range of aspect ratios. The following fit is obtained from this data

$$[\eta] = 1/(5A) + \lambda\left(1 + 0.058(A-1)^2/A - 0.029(\ln A)^2\right) + 4A^2/\left(5\ln(1+A^3)\right) \quad (3)$$

where $\lambda = 27/10 - 4/(5\ln 2) \approx 1.545844$. Note that this fit is exact in $A = 1$ (spheres) and in the limit $A \to \infty$ (needles). Cylindrical rods or stiff chains of connected beads behave similar[15] in that the intrinsic viscosity of rods also increases proportional to $[\eta] \sim A^2/\ln(A)$.

For dumbbells of identical spheres of diameter 1, placed at centre-to-centre distance $r_p$, the intrinsic viscosity increases faster than for ellipsoids. For two spheres at contact the intrinsic viscosity is the same as for ellipsoids of aspect ratio 2.7, and for $r_p \to \infty$ the limiting result is[13] $[\eta] \sim \frac{3}{4}r_p^2$. FIG. 1 shows that the intrinsic viscosity is quite sensitive to particle shape; any deviation from a sphere leads to an increase over $[\eta] = 2.5$. It is therefore not surprising that the coefficient of $\phi^2$ in the suspension viscosity (the Huggins coefficient $k_H$) is sensitive to particle-particle interactions, as attractive interactions may lead to aggregates, hence more extended objects. Analytical results are known for a few cases.[10] A recent estimate for hard spheres is[16] $k_H = 5.9147$, and for sticky hard spheres[10,17] $k_H \approx 21.4 - 12.2\Psi$, where $\Psi$ is a dimensionless osmotic second virial coefficient. Hence, $k_H$ increases by more than a factor 3 for a theta solvent ($\Psi=0$).

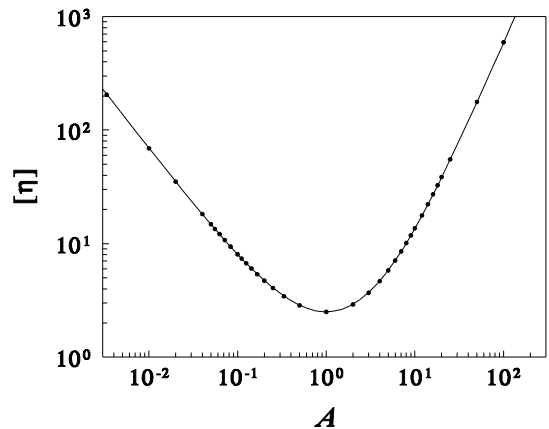

**FIG. 1.** Intrinsic viscosity for ellipsoids of revolution as function of aspect ratio $A$. The points are taken from Ref. 10; the fit curve is Eq. (3).

As a semi-empirical improvement to the Krieger-Dougherty equation, taking into account information on both $[\eta]$ and $k_H$, Martys[6] suggests $\eta = \eta_0(1-x)^{-2}[1+K_1x+K_2x^2]$, where $x = \phi/\phi_{max}$, $K_1 = [\eta]\phi_{max}-2$ and $K_2 = \phi_{max}^2k_H-2[\eta]\phi_{max}+1$. Note that Ref. 6 contains a printing error in the expression for $K_2$. In many cases the Huggins coefficient is not known. In those cases we may assume the Huggins coefficient predicted by the Krieger-Dougherty equation. Thus we can define a modified Krieger-Dougherty equation

$$\eta = \eta_0(1-x)^{-2}\left[1+(y-2)x+\frac{1}{2}(y-2)(y-1)x^2\right] \quad (4)$$

where $x = \phi/\phi_{max}$ and $y = [\eta]\phi_{max}$. In this modification the exponent by which the viscosity diverges corresponds to the established critical exponent for the electric conductivity of a random resistor network $t$, but the first and second coefficients of the volume fraction are retained. The exponent $t$ is obtained from the best estimate[18] for $t/\nu = 2.282\pm0.005$ and the correlation exponent[19] $\nu = 0.875\pm0.003$, which gives $t = 1.997 \pm0.008$. Thus, for practical purposes we may use the value $t = 2$.

Recent developments in the theory of fiber rheology are by Dhont and Briels[20] who developed a theory for the rheologic behaviour of rigid rods, including shear thinning. The intrinsic viscosity is a function of scaling parameter $L\phi/D$, where $L$ and $D$ are the rod length and diameter respectively. The theory compares well with experimental data for $L\phi/D < O(1)$. Zilman and Safran[21] studied associative rigid rods. These can form a cross-linked network that may show phase separation, just as was found beforehand for flexible associative polymers.[22] Moreover, Zilman and Safran[21] argue that *rigid* rods undergo a transition to a fiber bundle network that is driven by entropy. Finally, simulations by Bolhuis et al[23] on attractive spherocylinders show a rich phase diagram of demixing in isotropic, nematic and solid phases. This has strong implications for the rheology of fiber networks: both hydrodynamics and associative interactions need to be included.

## III. SIMULATION MODEL

The aim of this work is to define a simulation model that allows the largest possible coarse-graining of a colloidal dispersion, leading to hydrodynamics with least effort. In the spirit of Pryamitsyn and Ganesan[7] and of Pan et al.,[8] the simulation model therefore comprises two types of particles: the colloid particle (c) and the solvent particle (s). The function of the solvent particles is to generate long-range hydrodynamic interactions between the colloid particles. In this work solvent particles are ideal gas particles, whereas colloidal particles are modelled by (single) elastic spheres. How to impose the boundary conditions between solvent particles and colloid particles is described in section IV. Both solvent and colloid particles have noise and friction interactions with neighbouring particles up to a cut-off distance $r_c$. In the examples used in this work $r_c = 1.5$ is the same for all particles, but the forces are given are for arbitrary particle sizes.

### A. Particle interactions

The starting point for the present simulation model is the Fluid Particle Model (FPM) by Español.[24] Two types of interaction forces are used, conservative forces and dissipative forces. The dissipative forces consist of noise and friction. The friction force is implemented as a pair-wise friction between neighbouring particles, proportional to the relative velocity difference. It has been shown[25] that the friction should scale in a particular way to the particle sizes. For particles of unequal size the radial friction function is chosen as

$$\mathbf{F}_{ij}^R = -fR_{ij}\left(1 - h/R_{ij}\right)^2 \theta(1 - h/R_{ij})\,(\mathbf{v}_{ij} \cdot \mathbf{e}_{ij})\,\mathbf{e}_{ij} \quad (5)$$

where $h = r - D_{ij}$ is the gap width, $\theta(x)$ is the Heaviside step function and $\mathbf{e}_{ij} = (\mathbf{r}_i - \mathbf{r}_j)/|\mathbf{r}_i - \mathbf{r}_j|$; and we define the mean harmonic mean radius $R_{ij}$ and the mean particle diameter $D_{ij}$ as

$$R_{ij} = \frac{2R_i R_j}{R_i + R_j}; \quad D_{ij} = R_i + R_j \quad (6)$$

where $R_i$ is the radius of particle $i$. Note that for particles of the same size Eq. (5) is the standard DPD friction for a cut-off range $r_c = D_{ij} + R_{ij}$. In this formulation the friction force is $fR_{ij}$ for particles at contact.

The model further contains a shear friction proportional to the relative perpendicular velocities and spins. Similar models differing slightly from the original formulation by Español have been published

independently by Groot and Stoyanov[26] and by Pan et al[8]. We use

$$\mathbf{F}_{ij}^S = -\mu R_{ij}\left(1 - h/R_{ij}\right)^2 \theta(1 - h/R_{ij})\,[\,\mathbf{v}_{ij} - (\mathbf{v}_{ij} \cdot \mathbf{e}_{ij})\,\mathbf{e}_{ij} \\ + \mathbf{r}_{ij} \times (R_i\boldsymbol{\omega}_i + R_j\boldsymbol{\omega}_j)/D_{ij}\,] \quad (7)$$

where $\mathbf{r}_{ij} = \mathbf{r}_i - \mathbf{r}_j$, and where $\boldsymbol{\omega}_i$ is the spin of particle $i$. The combination of velocities and spins $[\,\mathbf{v}_{ij} - \ldots]$ appearing in Eq. (7) is in fact the difference between the particle velocities, plus an extra speed defined by the solid body rotation of each particle, extrapolated to the position of the other particle. $\mathbf{F}_{ij}^S$ is a force that points in a direction perpendicular to the line of contact between the particles. Brownian noise is introduced to maintain the desired temperature. In the original formulation of this model,[24] noise was introduced as components of symmetric, antisymmetric and traceless Wiener matrices. This is equivalent to the following (simpler) formulation

$$\mathbf{F}_{ij}^{Ran} = \sqrt{2R_{ij}kT}\left(1 - h/R_{ij}\right)\theta(1 - h/R_{ij}) \\ \left(\sqrt{f}\;\zeta_{ij}(t)\,\mathbf{e}_{ij} + \sqrt{\mu}\;\boldsymbol{\theta}_{ij}(t) \times \mathbf{e}_{ij}\right)/\sqrt{\delta t} \quad (8)$$

Here $\zeta_{ij}(t)$ is a random number of unit variance, $\boldsymbol{\theta}_{ij}(t)$ is a random vector with unit variance for *each* spatial component, and $\delta t$ is the time step taken. The outer product $\boldsymbol{\theta}_{ij} \times \mathbf{e}_{ij}$ guarantees a random force perpendicular to the line of contact. Note that the distance-dependence and the amplitude follow from the fluctuation-dissipation theorem.[8,24,26] For computational efficiency $\zeta_{ij}(t)$ and each component of $\boldsymbol{\theta}_{ij}(t)$ are drawn from a uniform distribution of the correct width, the results are indistinguishable from Gaussian random numbers. Total angular momentum is conserved by a pair-wise torque, given by

$$\boldsymbol{\tau}_i = -\frac{R_i}{D_{ij}}\sum_j \mathbf{r}_{ij} \times \mathbf{F}_{ij} \quad (9)$$

Finally, we integrate the equations of motion: $\dot{\mathbf{r}}_i = \mathbf{v}_i$; $\dot{\mathbf{v}}_i = \mathbf{F}_i / m_i$; $\dot{\boldsymbol{\omega}}_i = \boldsymbol{\tau}_i / I_i$, where $m_i$ and $I_i$ are the particle mass and moment of inertia.

The conservative interaction force is found from linear elasticity theory. Groot and Stoyanov[27] derived the repulsive force from the contact zone approximation as $F^{Rep} = (3/2)\,Eb(r_0 - r)$, where $r_0$ is the equilibrium distance between two fused spheres, $b$ is the radius of the neck and $E$ is proportional to the elastic modulus. For particles pressed together up to centre-to-centre distance $r$, the force derivative is thus given by

$$\partial F^{Rep}/\partial r = -\tfrac{3}{2}Eb \quad (10)$$

The elastic repulsion for two spheres of arbitrary size and distance is thus found by integrating the force derivative from the point of first contact down to centre-to-centre distance $r$. For this we need a relation between neck radius $b$ and distance $r$. From the geometry we find the relation $r = (R_i^2 - b^2)^{1/2} + (R_j^2 - b^2)^{1/2} \approx D_{ij} - b^2/R_{ij}$. After substitution and integration we find to a lowest order approximation

$$\mathbf{F}_{ij}^{Rep} = E\sqrt{R_{ij}}\left(D_{ij} - r\right)^{3/2}\mathbf{e}_{ij} \qquad (11)$$

This is the Hertz theory of elastic repulsion.[28] If $E$ is to be the elastic modulus of the material, the pre-factor should in fact be $2/(3(1-\nu^2))$ with $\nu$ the Poisson ratio. In the present formulation this factor is subsumed into the definition of $E$. This model is the starting point for the simulations described below. New elements will be introduced when and where appropriate.

## B. Semi-flexible fibers

An idealised model of very thin semi-flexible chains was proposed by Ramanathan and Morse.[29] Here, we want to include the hydrodynamic interactions. To model semi-flexible chains of particles *and* to maintain fluid boundary conditions at each particle, it is not sufficient to impose only bending interactions between two successive links. There are two fluid boundary conditions: the fluid flow velocity at the surface should vanish relative to surface motion both in radial direction and transverse direction. If we only implement a bending interaction between the links of three successive particles, the particles are still free to rotate relative to the underlying bond structure.

To fix both the bond structure of a particle cluster and the surface motion, each particle $i$ has three internal unit orientation vectors ($\mathbf{a}_i$, $\mathbf{b}_i$, $\mathbf{c}_i$). To evolve these vectors to the next time step, the particle spin $\boldsymbol{\omega}_i$ is integrated in time using a second order velocity-Verlet algorithm, together with the integration of particle positions.

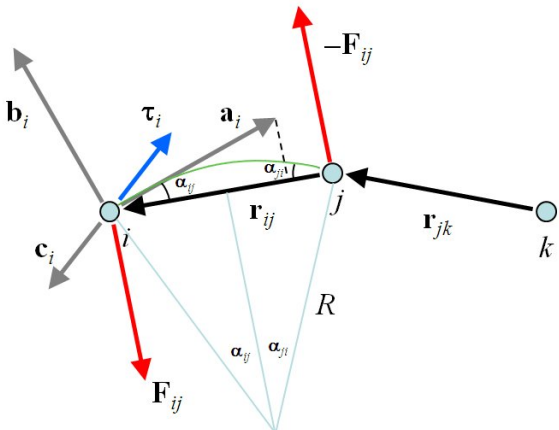

**FIG. 2** Frame of particle $i$ that is bound to particle $j$ (in grey); the resulting forces (in red) on particles $i$ and $j$; and the torque (in blue) on particle $i$. The green line shows the fiber of orientation $\mathbf{a}_i$ at point $i$.

The frame of particle $i$ that is bound to particle $j$, and the resulting forces and torque in this configuration are shown in FIG. 2. The particles $i$, $j$ and $k$ form a semi-flexible fibre with stiffness $K_b$. To maintain a linear conformation and to maintain a fixed orientation of particle $i$ relative to the chain we define the director of the fiber in point $i$ by the unit vector $\mathbf{a}_i$ and impose a force to align it with the coordinate difference $\mathbf{r}_{ij} = \mathbf{r}_i - \mathbf{r}_j$. The green line in FIG. 2 shows the fiber as a (stiff) beam between particles $i$ and $j$, with local orientations $\mathbf{a}_i$ and $\mathbf{a}_j$ that are misaligned with vector $\mathbf{r}_{ij}$ by angles $\alpha_{ij}$ and $\alpha_{ji}$ respectively. If this beam is fixed at points $i$ and $j$, a bending moment is exerted, given by $\tau = K_b/R$, where $R$ is the radius of curvature (see FIG. 2). The bending force exerted in point $i$ is thus obtained as $F = \tau/r_{ij} = K_b/Rr_{ij}$. The local curvature is $1/R = 2\sin(\alpha_{ij})/r_{ij}$, hence the total force acting on particle $i$, needed to bend the beam is $F = 2K_b\sin(\alpha_{ij})/r_{ij}^2$. To code this efficiently we calculate the perpendicular component of $\mathbf{a}_i$ to the unit vector $\mathbf{e}_{ij} = \mathbf{r}_{ij}/|\mathbf{r}_{ij}|$ and define the force

$$\mathbf{F}_{ij}^b = -K_b \sin(\alpha_{ij})\,\mathbf{e}_\alpha\,/\,r_{ij}^2 = -K_b(\mathbf{a}_i - \mathbf{e}_{ij}(\mathbf{a}_i \cdot \mathbf{e}_{ij}))\,/\,r_{ij}^2 \quad (12)$$

where $\mathbf{e}_\alpha$ is a unit vector in the direction of the deflection. To align the orientation of the vector $\mathbf{r}_{ij}$ with the frame of particle $i$ the force $\mathbf{F}_{ij}^b$ is exerted on particle $i$ and the opposite force $-\mathbf{F}_{ij}^b$ is exerted on particle $j$. Finally, to conserve angular momentum an opposite torque acts on particle $i$, as shown in FIG. 2:

$$\boldsymbol{\tau}_i = -\mathbf{r}_{ij} \times \mathbf{F}_{ij}^b \qquad (13)$$

Note that this torque acts only on particle $i$, since it is this particle that is being aligned with the bond vector $\mathbf{r}_{ij}$. A similar force and torque is defined for particle $j$ (with indices $ij$ in Eq. (12) replaced by $ji$), where force $\mathbf{F}_{ji}^b$ is calculated for particle $j$, and an opposite force acts on particle $i$. For each linked pair $ij$ the forces and torques defined in Eqs. (12) and (13) are applied. The result is that all bonds in a chain are forced to align, and the internal frames of the particles are aligned with the bond vectors.

So far we have only discussed a bending stiffness interaction along the chain, but the particles are still free to rotate along an axis parallel to the chain. To prevent this rotation we use the other internal vectors of the particles. If two particles $i$ and $j$ are linked, we require that the vectors $\mathbf{b}_i$ and $\mathbf{b}_j$ align. However, a misalignment where e.g. $\mathbf{b}_j$ is tilted in the $\mathbf{r}_{ij}$ direction is in fact a bending deformation. To prevent double counting of the bending forces and deal with torsion only, we first rotate the frame of particle $i$ to that of $j$, to undo the bending. Thus, $\mathbf{b}_i$ is rotated around the axis $\mathbf{u} = (\mathbf{a}_i \times \mathbf{a}_j)/|\mathbf{a}_i \times \mathbf{a}_j|$ to a vector $\mathbf{b}'_i$ to undo the bending. Since parallel vector components to $\mathbf{u}$ are invariant to rotation around the $\mathbf{u}$ axis, and since ($\mathbf{a}_i$, $\mathbf{b}_i$, $\mathbf{c}_i$) forms an orthonormal frame, the outer product $\mathbf{b}'_i \times \mathbf{b}_j$ is obtained conveniently by projecting $\mathbf{b}_i$, $\mathbf{c}_i$, $\mathbf{b}_j$ and $\mathbf{c}_j$ onto $\mathbf{u}$. We thus obtain the torques and forces

$$\begin{aligned}
\boldsymbol{\tau}_i &= K_t\ (\mathbf{b}'_i \times \mathbf{b}_j) \cdot \mathbf{a}_i\ (\mathbf{a}_i\,/\,r_{ij}) \\
&= K_t[(\mathbf{c}_i \cdot \mathbf{u})(\mathbf{b}_j \cdot \mathbf{u}) - (\mathbf{b}_i \cdot \mathbf{u})(\mathbf{c}_j \cdot \mathbf{u})]\,\mathbf{a}_i\,/\,r_{ij} \\
\boldsymbol{\tau}_j &= -K_t(\mathbf{b}'_i \times \mathbf{b}_j) \cdot \mathbf{a}_j\ (\mathbf{a}_j\,/\,r_{ij}) \\
\mathbf{F}_i &= -\mathbf{F}_j = \mathbf{r}_{ij} \times (\boldsymbol{\tau}_i + \boldsymbol{\tau}_j)\,/\,r_{ij}^2
\end{aligned} \qquad (14)$$

where $K_t$ is the torsional stiffness. For simplicity we take $K_t = K_b$. Because $\mathbf{a}_i$ and $\mathbf{a}_j$ can be misaligned by bending, the forces given in Eq. (14) are needed to balance the sum of torques.

The frame vectors $\mathbf{a}_i$, $\mathbf{b}_i$, and $\mathbf{c}_i$ are updated using the algorithm

$$\mathbf{a}_i(t+\delta t) := \mathbf{a}_i(t) + \delta t \boldsymbol{\omega}_i(t) \times \mathbf{a}_i(t) + \tfrac{1}{2}(\delta t)^2 \boldsymbol{\tau}_i(t) \times \mathbf{a}_i(t) \tag{15}$$

Next, normality is restored via

$$\mathbf{a}_i' := 2\mathbf{a}_i /\left(1 + \mathbf{a}_i \cdot \mathbf{a}_i\right) \tag{16}$$

Note that if the square length of a frame vector deviates slightly from 1, $\mathbf{a}_i \cdot \mathbf{a}_i = 1+\varepsilon$, the new vector has square length $\mathbf{a}_i' \cdot \mathbf{a}_i' = 1-\varepsilon^2/4$. Thus, small errors in the frame vector length, picked up in the integration procedure, are quickly removed and we obtain a very stable algorithm. Finally, orthonormality is restored by applying

$$\mathbf{a}_i' := \left(\mathbf{a}_i - (\mathbf{a}_i \cdot \mathbf{b}_i)\mathbf{b}_i\right)/\left(1 - \tfrac{1}{2}(\mathbf{a}_i \cdot \mathbf{b}_i)^2\right) \tag{17}$$

for the combinations $(\mathbf{a}_i, \mathbf{b}_i)$; $(\mathbf{b}_i, \mathbf{c}_i)$; and $(\mathbf{c}_i, \mathbf{a}_i)$. Again, this restores orthonormality to order $\varepsilon^2$ yet avoids taking square roots.

## C. Measuring viscosity

To measure viscosity in simulation two methods can be employed. The first is to impose a constant shear rate using the Lees-Edwards boundary conditions.[30] In this method a velocity in the $x$-direction is added or subtracted to a particle velocity when it moves through the upper or lower boundary of the simulation box. Under these boundary conditions the shear stress is calculated. In general, the stress tensor has two contributions, a kinetic term and an interaction term, and is given by

$$\sigma^{\alpha\beta} = -\frac{1}{V}\sum_i m_i v_i^\alpha v_i^\beta - \frac{1}{V}\sum_{i>j} F_{ij}^\alpha r_{ij}^\beta \tag{18}$$

The viscosity is obtained by dividing the shear stress by the shear rate, i.e.

$$\eta = \sigma^{xy}/\dot{\gamma} \tag{19}$$

Note that the first term in Eq. (18) (kinetic momentum transport) is non-zero in shear because particles arriving at the point of measurement from below have a negative average $x$-velocity and those arriving from above a positive, for positive shear rate. The resulting kinetic viscosity is proportional to the diffusion constant,[31] $\eta_K = \rho D/2$. If we work at low temperature, or high viscosity $D \to 0$, we can neglect this term.

At low viscosity we cannot ignore the kinetic contribution. In that case we use the method proposed by Backer et al.[32] In this method an upward gravity force is exerted on all particles at the left half of the system, and an equal downward gravity force on all particles at the right half of the system. This produces two parabolic (Poisseuille flow) velocity profiles in opposite direction. Let the size in the $x$-direction be $L$, and the force on each particle be $\pm g$, depending on the position in the system, then the velocity profile will be proportional to

$$\Lambda(x) = \begin{cases} x(L/2-x) & (\text{if } x < L/2) \\ (L-x)(L/2-x) & (\text{if } x > L/2) \end{cases} \tag{20}$$

The fluid viscosity thus follows as

$$\eta = \frac{\rho g L^4}{960\langle\Lambda(x)v_z(x)\rangle} \tag{21}$$

where $<A> = \int A dx/L$ is an average of quantity $A$ over the simulation box, and $\Lambda(x)$ is the polynomial defined in Eq (20).

In some cases it may be convenient to work at a low temperature and set e.g. $kT = 10^{-3}$ to simulate high viscosity fluids. Identical physics is obtained if we simultaneously rescale temperature, elasticity modulus, friction coefficient and time as follows:

$$kT \to 1; \quad E \to E^* = E/kT;$$
$$f \to f^* = f/\sqrt{kT}; \quad t \to t^* = t \cdot \sqrt{kT} \tag{22}$$

## IV. MODIFIED FRICTION FORCES AND STICK BOUNDARY CONDITIONS

To simulate the hydrodynamic interaction between colloidal particles two particle types are used in simulation, the colloid particle (c) and the solvent particle (s). The function of the solvent particle is to generate long-range hydrodynamic interactions between the colloid particles. This can be done in two ways, 1) by using modified friction forces, and 2) by imposing the correct the particle-fluid boundary conditions.

A third option has been investigated, the free draining approximation, where the repulsive force between colloid and solvent vanishes, $E_{cs} = 0$; no colloid-solvent boundary is present. The solvent acts as a heat bath; there is *only* friction and noise interaction between colloid and solvent. If the colloidal particles repel each other with elasticity $E_{cc}/kT = 10^6$ and the solvent is an ideal gas ($E_{ss} = 0$), viscosity diverges close to dense random packing. However, for low colloid volume faction we find $\eta \approx \eta_0 \exp(\alpha\phi)$, where $\alpha$ is a constant. For a solvent without particles we find at low temperature ($kT = 0.001$) $\eta \propto \rho^{2.72\pm0.03}$ where $\rho$ the concentration of solvent particles. Thus, if we add a (low) concentration $\phi/V_c$ of colloid particles, viscosity must follow $\eta \propto (\rho+\phi/V_c)^{2.72} \approx \rho^{2.72}\exp(2.72\phi/V_c\rho)$. The pre-factor of $\phi$ in the exponential can take on any desired value $[\eta] = \alpha = 2.72/V_c\rho$ by choosing the right solvent concentration.

Even though this leads to the correct intrinsic viscosity by construction, it does so for the wrong reason. The nature of the free draining approximation is that viscosity is increased simply because more friction centres are added to the system when colloid particles are included, and not because of the flow boundary conditions at solid objects. This means that we cannot rely upon this model to predict the shear viscosity for any shape other than spheres. For this reason the free draining approximation is discarded.

To demonstrate the failure of the free draining approximation explicitly some simulations were done to obtain the viscosity using the Backer method[32] for a pure solvent of 5100 ideal solvent particles (s) in a box of size 40×14×14, and the same solvent with 748 colloidal particles (c) added. Three cases were considered 1) all particles moved freely, 2) the colloid particles were grouped in 374 dimers, and 3) the colloid particles were grouped in 187 linear tetramers. The friction used was $f = \mu = 1$, and the particle repulsions were $E_{cc} = 1000$ and $E_{ss} = E_{cs} = 0$. In all cases the colloid volume fraction is exactly $\phi = 0.05$. At high temperature ($kT = 1$) we find viscosity ratios $\eta/\eta_0 = 1.105\pm0.005$, $1.105\pm0.005$ and $1.174\pm0.005$ respectively for cases 1, 2 and 3. Since this ratio equals $\exp(\alpha\phi)$, we obtain $\alpha = 2.0\pm0.1$, $2.0\pm0.1$, and $3.2\pm0.1$ for the three cases respectively. Since $\alpha$ should be *proportional* to the intrinsic viscosity, and since the exact values are known for monomers ($[\eta] = 2.5$), touching spheres ($[\eta] = 3.45$) and ellipsoids of aspect ratio 4 ($[\eta] = 4.66$) (see section II), it is clear that the simulation result is wrong. Even if we multiply the values of $\alpha$ by 1.25 to match the monomer result, we still find 2.5 for dimers and 4.0 for tetramers, i.e. too low. If the same simulations are repeated at low temperature ($kT = 0.001$) quite different results are obtained ($\eta/\eta_0 = 1.531\pm0.006$, $1.64\pm0.01$ and $1.84\pm0.01$ respectively), leading to $\alpha = 8.5\pm0.1$, $9.9\pm0.2$, and $12.2\pm0.4$ respectively. These values are equally wrong. This shows that other shapes than spheres, such as dimers and rods formed by spheres, are not simulated well in the free draining approximation.

## A. Applying modified friction forces

One approximate way of including the boundary conditions is to choose the shear and radial friction forces between colloid and solvent such that the right hydrodynamic forces are exerted on a particle. This is the strategy taken by Pan *et al.*[9] The effect of hydrodynamic interactions is that the friction force on a moving particle in a stationary fluid is given by

$$\mathbf{F} = -6\pi\eta R \mathbf{v} \qquad (23)$$

where $\eta$ is the fluid viscosity and $\mathbf{v}$ the velocity of the particle. Similarly, a rotating sphere in an infinite fluid experiences a torque given by

$$\boldsymbol{\tau} = -8\pi\eta R^3 \boldsymbol{\omega} \qquad (24)$$

where $\boldsymbol{\omega}$ is the particle spin. If only radial friction is employed[7] these relations are not satisfied and consequently the absolute simulated volume fraction does not correspond to the real colloid volume fraction. To repair this flaw, Pan *et al*[8] introduced a modified FPM with radial and shear frictions between colloid and solvent particles, different from the values used between solvent particles mutually, so as to satisfy Eq. (23) and (24). This approximates the true boundary conditions at the particle surface by imposing the correct forces.

However, for a practical case where colloid rheology is simulated,[9] very small time steps need to be taken. Moreover, since the parameterisation procedure was done numerically, this method can only be applied to a single colloid size and not to a more realistic case of a distribution of sizes. For these reasons the implementation of the stick boundary condition has been reanalysed.

Because the force and torque given in Eq. (23) and (24) are simple power laws of the particle radius, friction cannot depend on the fluid radius of friction $r_c$. In particular, it cannot be a function of the mean harmonic radius like in Eq. (5) or (7). For this reason we have to impose new rules for the interaction between colloidal particles and solvent particles. The simplest adaptation that satisfies this requirement is to treat the solvent particle in colloid-solvent interactions as a point particle, and to replace the mean harmonic radius by the colloid radius. Thus, the radial and friction forces between colloid and solvent are replaced by

$$\mathbf{F}_{ij}^R = -f^{CS} R_i \left(p - r_{ij}/R_i\right)^2 \theta(p - r_{ij}/R_i) \left(\mathbf{v}_{ij} \cdot \mathbf{e}_{ij}\right) \mathbf{e}_{ij}$$

$$\mathbf{F}_{ij}^S = -\mu^{CS} R_i \left(p - r_{ij}/R_i\right)^2 \theta(p - r_{ij}/R_i) \qquad (25)$$

$$\left[\mathbf{v}_{ij} - \left(\mathbf{v}_{ij} \cdot \mathbf{e}_{ij}\right) \mathbf{e}_{ij} + r_{ij} \times \boldsymbol{\omega}_i\right]$$

where $i$ is the colloidal particle and $j$ is the fluid particle. Parameter $p$ sets the range of the interaction. If we choose $p = 3$ the friction interaction extends one particle diameter away from its surface, i.e. up to $r_{ij} = 1.5D_i$. This is a practical choice, since the maximum particle diameter is $D_i = 1$, which means that also for the colloid-colloid friction and the fluid-fluid friction neighbours are searched for up to the cut-off distance $r_c = 1.5$.

On top of the radial and shear friction forces, we impose an elastic repulsion between colloid particles and fluid particles to simulate an elastic core interaction at $r_{ij} = 0.5$ (Eq. (11) using $D_{ij} = R_{ij} = R_i = 0.5$). To estimate the values of the radial and shear friction factors, and to obtain their scaling with particle size, we use the known solution of the flow field around a moving or rotating particle[33] and *assume* that this flow field is not disturbed by the friction forces, but is the same as for a particle with true stick boundary conditions. This is an approximation, because what is essentially a boundary effect is smeared out over a finite volume by the friction forces. The approximation is expected to be reasonable at low Reynolds numbers and small friction. In Appendix A the torque on a sphere is calculated for the force field of Eq. (25). When this is equated to the theoretical value Eq. (24), we find the shear friction factor as

$$\mu^{CS} = 0.18885 \frac{\eta}{\rho R_i^3} \qquad (26)$$

where $\rho$ is the density of solvent particles and $\eta$ is the fluid viscosity. The numerical value is given for the range $p = 3$ that is used here. The Appendix gives the value for arbitrary force range.

In a similar way we find the friction force of a moving particle (for $p = 3$) to match the theoretical result Eq. (23), if the radial friction coefficient is chosen as

$$f^{CS} = 2.52580 \frac{\eta}{\rho R_i^3} \qquad (27)$$

It should be noted that both the calculated shear friction coefficient and the radial friction coefficient strongly depend on the value of the force cut-off range.

These relations were tested as follows. Firstly, the parameters $kT = 0.001$, $f = 1$, $\mu = 0$ and a liquid density $\rho r_c^3 = 3$, (i.e. $\rho = 0.8888$) were selected. The viscosity $\eta = 0.03751(4)$ was determined using the Backer method[32] using three runs of $10^5$ steps ($\delta t = 0.01$). Using Eqs. (26) and (27) the radial and shear frictions should be $f^{CS} = 0.852$ and $\mu^{CS} = 0.064$ for colloid radius $R_i = 0.5$. To test this relation a single particle with fixed spin $\omega = 0.1$ was inserted, maintaining a constant solvent density. According to Eq. (24) we expect a torque $\tau = -0.012$. Over three runs of $10^6$ steps we actually found $\tau = -0.0117(2)$. This confirms Eq. (26).

To test Eq. (27) the colloid particle was moved by a constant velocity with $v = 0.1$ and the friction force was measured. Eq. (23) predicts $F = -0.0354$. A first run gave $F = -0.059$, i.e. too high by a factor 1.7. However, in these conditions the Peclet number $Pe = 6\pi\eta R_i^2 v/kT \approx 18$, and the Reynolds number $Re = 2\rho v R_i/\eta \approx 2.4$. It can be expected that Eqs. (26) and (27) break down at high Reynolds numbers because the flow field used in Appendix A is the solution in the limit $Re \rightarrow 0$, and to break down at high Peclet numbers because the solvent is compressible. Therefore the velocity was brought back to $v = 0.01$; in a run of $10^6$ steps a force of $F = -0.0044$ was obtained, i.e. some 24% too high which is not too bad given the approximations.

To simulate more accurately at low Peclet number, parameters were changed to $kT = 1$, $f = 1$ and $\mu = 0$. Note that this alters the time scale and viscosity according to the scaling laws given in Eq. (22). We obtain a viscosity $\eta = 0.48(1)$, hence the colloid friction parameters should be $f^{CS} = 11$ and $\mu^{CS} = 0.82$. For a linear velocity $v = 0.1$ and spin $\omega = 0.1$ we thus expect $F = -0.454$ and $\tau = -0.151$.

In actual simulations at time step $\delta t = 0.04$, simulating just one colloidal particle in a bulk fluid of density $\rho r_c^3 = 3$, we find that the fluid density drops significantly towards the particle-fluid surface, and the pair correlation shows too much overlap. Even at step $\delta t = 0.01$ the pair correlation is incorrect. Only at $\delta t = 0.005$ the fluid density near the particle surface is homogeneous, but due to the small step size long runs are necessary. Over two runs of $10^6$ steps each we find $F = -0.448\pm0.018$ and of $\tau = -0.114\pm0.009$. In this case the linear force is correct, but the simulated torque is some 25% too low.

The ratio of the two friction factors always satisfies $f^{CS}/\mu^{CS} > 10.8$ (see Appendix A). For the friction range parameter $p = 3$ this ratio is 13.4. This suggests that in practice the radial boundary condition is more important than the transverse boundary condition. Whereas the shear friction is quite moderate, the radial friction is more than an order of magnitude higher than between fluid particles. This indicates that the main reason to use a small time step in the Pan-Pivkin-Karniadakis method[8] to simulate stick boundary conditions is high friction in radial direction. We thus face the problem that we need a low viscosity liquid to simulate at low Peclet number, but a high (or infinite) viscosity at the particle-fluid boundary.

This problem may be solved by using another integrator for the viscous interaction, e.g. based on Shardlow's splitting method S1,[34] or by using the Lowe-Anderson thermostat.[35] The analysis by Nikunen et al.[36] points at these two methods as the most promising integrators. For the ideal gas both methods perform superior to other methods, but for fluids with repulsion velocity-Verlet is better than S1. To my knowledge both methods have been used only for radial friction, and require adaptation to account for the transverse friction modes. But even if the small time step problem would be solved, this would still solve only an approximate model if we apply it to the friction forces of Eqs. (25-27). As pointed out above, the analytic model is valid only at low Reynolds numbers, and even there it is an approximation because what is essentially a surface effect is smeared out over a larger area.

The preferred approach would be to impose the boundary conditions exactly. A method to impose stick boundary conditions in DPD or Smooth Particle Hydrodynamics has been described by several authors,[37,38] and has been used to simulate colloidal particles and polymers.[6,39] However, this essentially rests on a discrete representation of the colloid particles, where a number of (DPD) particles is frozen to form a colloid particle.[6,39] The DPD particles inside a colloidal particle are then given a (virtual) velocity opposite to a colliding solvent particle. It is obvious that this method requires large numbers of DPD particles to have sufficient resolution for the colloidal particle, which limits the accessible simulation size. In a minimal representation of the problem, the colloidal particles should be represented by a *single* simulation unit, a geometrical hard sphere.

To impose stick boundary condition at such hard spheres, the velocity difference of solvent particle and colloid particle should be randomized when a solvent particle hits a hard sphere surface. Velocity randomization is exactly what is done by the Lowe-Anderson thermostat; it is a Monte Carlo sampling scheme for the particle velocities. The difficulty is to combine this with the DPD or FPM method for the fluid away from particle surfaces, and to sample in such a way that momentum and angular momentum are conserved. This is described in the next section.

## B. Applying stick boundary conditions

There are two boundary conditions to be satisfied when a solvent particle collides with a colloidal particle: the *mean* radial velocity should be zero relative to the surface, and the *mean* tangential velocity should vanish. This can be achieved by two collision rules. First the radial collision rule is discussed. A crucial step in applying stick boundary conditions is to combine the Lowe-Anderson thermostat with the DPD or FPM method to simulate a relatively low viscosity fluid. This step was taken by Stoyanov and Groot.[40] In this simulation method all particles are thermostated through interaction with their neighbours. However a choice is made between *either* soft (DPD or FPM) noise and friction, *or* a Monte Carlo step to exchange momentum and replace the radial velocity difference by a new value taken from a Gaussian distribution. In the Monte Carlo step two interacting particles $i$ and $j$ are given new velocities according to

$$\begin{cases} \mathbf{v}_i' = \mathbf{v}_i + \Delta\mathbf{p}/m_i \\ \mathbf{v}_j' = \mathbf{v}_j - \Delta\mathbf{p}/m_j \end{cases} \quad (28)$$

The momentum exchange $\Delta\mathbf{p}$ in the Lowe-Anderson thermostat is given by

$$\Delta\mathbf{p} = M_{ij}[\xi\sqrt{kT/M_{ij}} - (\mathbf{v}_{ij} \cdot \mathbf{e}_{ij})]\,\mathbf{e}_{ij} \quad (29)$$

where $\xi$ is a standard normally distributed random variable, and where $m_i$ and $m_j$ are particle masses, $M_{ij} = m_i m_j/(m_i+m_j)$, $\mathbf{e}_{ij} = (\mathbf{r}_i-\mathbf{r}_j)/r_{ij}$ and $\mathbf{v}_{ij} = \mathbf{v}_i-\mathbf{v}_j$.

In the present application, we wish to apply this thermostat only to overlapping particle pairs, and replace the radial velocity by a *positive* value taken from a Gaussian distribution, i.e. we replace $\xi$ in Eq. (29) by $|\xi|$. This serves as a hard core interaction; particles are reflected back when they overlap. However, in doing so it was observed that this procedure leads to about 1.5% artificial temperature reduction in a simulation of 500 colloidal particles and 638 fluid particles. The reason for this is that fast solvent particles have a higher probability of crossing the boundary of a colloidal particle, since these take larger steps. The tail of the velocity distribution is therefore overrepresented in collisions. If these velocities are replaced by a number from a standard Gaussian distribution, the system is cooled artificially.

To generate the *correct* radial velocity distribution we simply reflect the overlapping particles back. Thus, we will use Eq. (25) for the radial friction with $f^{CS} = f$, when the solvent particle does not overlap the hard core of the colloid particle. However, for overlapping pairs of colloidal particles $i$ and fluid particles $j$ (i.e. with centre-to-centre distance $r_{ij} < R_i$) the pair $ij$ is stored in a neighbour list. After the main force loop and velocity update all particles on the neighbour list exchange momentum as in Eq. (28), but now we use the rule

$$\begin{cases} \Delta\mathbf{p} = -2M_{ij}(\mathbf{v}_{ij} \cdot \mathbf{e}_{ij})\,\mathbf{e}_{ij} & \text{if } (\mathbf{v}_{ij} \cdot \mathbf{e}_{ij}) < 0 \\ \Delta\mathbf{p} = 0 & \text{if } (\mathbf{v}_{ij} \cdot \mathbf{e}_{ij}) > 0 \end{cases} \quad (30)$$

This effectively simulates a hard core interaction because the relative momentum changes sign. Note that sign reversal is only applied when the relative velocity is negative; for positive velocities the overlapping fluid particle is already "on its way out". The contribution to the pressure tensor of a collision is given by

$$\delta P^{\alpha\beta} = -2M_{ij}(\mathbf{v}_{ij} \cdot \mathbf{e}_{ij})\,e_{ij}^{\alpha}\,r_{ij}^{\beta}/V \quad (31)$$

where $\mathbf{v}_{ij}$ is the velocity difference *prior* to collision, $\alpha,\beta$ denote spatial directions and $V$ is the system volume.

A minor correction to this procedure is apply a velocity sign reversal (i.e. a collision) not only to overlapping particles, but also to some particle pairs close to overlap. Thus, a collision step is applied to colloid-fluid particle pairs with a probability $P(r_{ij})$, given by

$$P(r_{ij}) = \begin{cases} 1 & \text{if } r_{ij} < R_i \\ 1-(r_{ij}-R_i)/s & \text{if } R_i < r_{ij} < R_i + s \\ 0 & \text{if } r_{ij} > R_i + s \end{cases} \quad (32)$$

where $R_i$ is the radius of the colloidal particle. The range $s$ is determined such that collision *on average* takes place at the particle radius. Since the overlap distance is proportional to the time step, $s \propto \delta t$. For a range of time steps $s$ was varied, and the average $\langle r_{ij}-R_i\rangle$ was determined over all collisions. Up to time step $\delta t = 0.025$ this is found to vanish at $s = 1.691\delta r\sqrt{kT} + 0.32\,(\delta t)^2 kT$.

With the collision rule Eq. (30) for the radial velocity, and the shear friction Eq. (25) with the coefficient in Eq. (26) to impose the tangential boundary condition, we already have a large improvement over the modified FPM method, because the main reason why small time steps need to be taken is the strong radial friction. Even if the viscosity is not determined very accurately, this still leads to a reasonable result because the shear forces are responsible for some 10% of the viscosity. The collision rule for the radial boundary condition ensures the correct location of the boundary condition, and it allows for much larger time steps than with noise and friction forces over a range $p \sim 1.01$ (see Eq (25)) as used by Pan *et al.*[9] Indeed, for a small range $p \sim 1+\epsilon$ the friction forces at particle contact diverge as $f^{CS}\epsilon \sim \epsilon^{-3}$ and $\mu^{CS}\epsilon^2 \sim \epsilon^{-2}$.

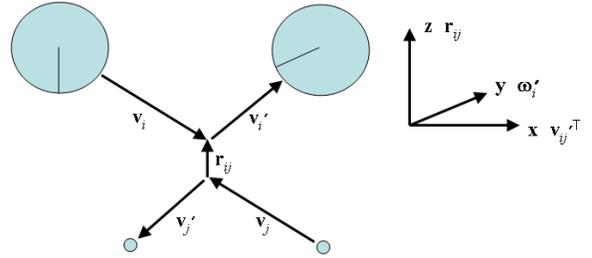

**FIG. 3** Collision between colloid particle $i$ and fluid particle $j$. Primed symbols refer to quantities are after collision.

However, the best option would be to impose the tangential boundary condition via a collision rule at the particle surface. This does not rely on an effective force field to impose the correct forces and torques on particles, and uses the actual physical mechanism that causes viscosity increase in a colloid suspension. To date, no algorithm has been published to generalise the Lowe-Anderson thermostat to a sampling algorithm for both radial and tangential velocities. Spin and velocity are both Gaussian distributed, yet changes should be correlated by conservation of angular momentum. This makes the development of such a general Monte Carlo scheme more challenging.

To tackle this problem, consider a colloidal particle $i$ and a fluid particle $j$, with initial velocities $\mathbf{v}_i$ and $\mathbf{v}_j$, and let $i$ have initial spin $\boldsymbol{\omega}_i$. Let the transverse velocity $\mathbf{v}_{ij}^T = \mathbf{v}_{ij} - (\mathbf{v}_{ij} \cdot \mathbf{e}_{ij}) \, \mathbf{e}_{ij}$ for the sake of the argument point in the $x$-direction and let the centre-to-centre vector at collision $\mathbf{r}_{ij}$ point in the $z$-direction (see FIG. 3). Stick boundary conditions now imply that the spin after collision will point in the $y$-direction (see FIG. 3). In particular, stick conditions imply that after collision we have

$$\mathbf{v}_{ij}^{\prime T} = -\mathbf{r}_{ij} \times \boldsymbol{\omega}_i' \qquad (33)$$

Note that superscript $T$ denotes transverse components. In general, we have conservation of momentum, so in any collision Eq. (28) holds. Since the force on particle $i$ acts on its surface, its angular momentum changes by $-\mathbf{r}_{ij} \times \Delta \mathbf{p}$, thus its spin changes according to

$$\boldsymbol{\omega}_i' = \boldsymbol{\omega}_i - \mathbf{r}_{ij} \times \Delta \mathbf{p} / I_i \qquad (34)$$

Substitution into Eq. (33) eliminates $\boldsymbol{\omega}_i'$ on behalf of the spin before collision:

$$\begin{aligned} \mathbf{v}_{ij}^{\prime T} &= -\mathbf{r}_{ij} \times [\boldsymbol{\omega}_i - \mathbf{r}_{ij} \times \Delta \mathbf{p} / I_i] \\ &= -\mathbf{r}_{ij} \times \boldsymbol{\omega}_i + \mathbf{r}_{ij} \times [\mathbf{r}_{ij} \times \Delta \mathbf{p}] / I_i \end{aligned} \qquad (35)$$

Next, we substitute the collision rule Eq. (28) to eliminate the unknown new velocity, and find

$$\begin{aligned} (\mathbf{v}_i + \Delta \mathbf{p} / m_i - \mathbf{v}_j + \Delta \mathbf{p} / m_j)^T &= -\mathbf{r}_{ij} \times \boldsymbol{\omega}_i + \mathbf{r}_{ij} \times [\mathbf{r}_{ij} \times \Delta \mathbf{p}] / I_i \\ \Rightarrow \qquad \mathbf{v}_{ij}^T + \Delta \mathbf{p} / M_{ij} &= -\mathbf{r}_{ij} \times \boldsymbol{\omega}_i + \mathbf{r}_{ij} \times [\mathbf{r}_{ij} \times \Delta \mathbf{p}] / I_i \\ \Rightarrow \Delta \mathbf{p} / M_{ij} - \mathbf{r}_{ij} \times [\mathbf{r}_{ij} \times \Delta \mathbf{p}] / I_i &= -(\mathbf{v}_{ij}^T + \mathbf{r}_{ij} \times \boldsymbol{\omega}_i) \end{aligned} \qquad (36)$$

Note that $\Delta \mathbf{p}$ is by definition perpendicular to $\mathbf{r}_{ij}$ as here we consider only the transversal momentum change. In general $\mathbf{r} \times [\mathbf{r} \times \Delta \mathbf{p}] = -r^2 \Delta \mathbf{p} + \mathbf{r} \, (\mathbf{r} \cdot \Delta \mathbf{p})$, but since $\Delta \mathbf{p}$ is perpendicular to $\mathbf{r}_{ij}$ the second term does not contribute. Thus we solve the momentum change as

$$\Delta \mathbf{p} = -\frac{(\mathbf{v}_{ij}^T + \mathbf{r}_{ij} \times \boldsymbol{\omega}_i)}{1 / M_{ij} + r_{ij}^2 / I_i} \qquad (37)$$

Substitution into Eq. (28) and (34) gives the full collision rule. The remarkable result is that the inverse effective mass $1/M_{ij} = 1/m_i + 1/m_j$ is replaced by a new inverse effective mass given by

$$1 / \mu_{ij} = 1 / m_i + 1 / m_j + r_{ij}^2 / I_i \qquad (38)$$

i.e. the moment of inertia starts to play the role of mass.

At this point it is necessary to consider the energy change of a collision. By straightforward substitution of

Eq. (37) into Eq. (28) and (34) we find that there is *always* an energy loss given by

$$\Delta E = -\tfrac{1}{2} \Delta \mathbf{p}^2 / \mu_{ij} \qquad (39)$$

i.e. we loose the kinetic energy of a hypothetic particle of mass $\mu_{ij}$ and momentum $\Delta \mathbf{p}$. To correct for this energy loss, we have to add to $\Delta \mathbf{p}$ a noise term that corresponds exactly to this "lost particle". Since this noise is uncorrelated to the velocities and spin, stick boundary conditions will be satisfied *on average*, but the energy corresponding to this noise adds up to the total energy change, which then balances out to zero *on average*. Thus, we modify the momentum change Eq. (37), and arrive at our final result for the collision rule:

$$\begin{aligned} \Delta \mathbf{p} &= -\mu_{ij} \left[ \mathbf{v}_{ij} - \mathbf{e}_{ij}(\mathbf{e}_{ij} \cdot \mathbf{v}_{ij}) + \mathbf{r}_{ij} \times \boldsymbol{\omega}_i + \boldsymbol{\theta}_{ij} \times \mathbf{e}_{ij} \sqrt{kT / \mu_{ij}} \right] \\ \Delta \boldsymbol{\omega}_i &= -\mathbf{r}_{ij} \times \Delta \mathbf{p} / I_i \end{aligned} \qquad (40)$$

where the effective mass $\mu_{ij}$ is defined in Eq. (38). The first two terms in Eq. (40) define the transverse velocity, the third includes the velocity effect of solid body rotation of particle $i$ at separation vector $\mathbf{r}_{ij}$, and $\boldsymbol{\theta}_{ij}$ is a 3D Gaussian random vector of unit variance in *each* direction. The outer product produces a vector normal to $\mathbf{r}_{ij}$.

It was shown by Stoyanov and Groot[40] that, in the presence of friction forces, the velocity integration should not be mixed with a Monte Carlo update for the velocity. Similarly, the present collision rule should be applied after the velocity update. The correct order of procedures is:

- update particle positions
- update velocities by a half step,
- put particles in cells
- search neighbours, calculate neighbour forces and place overlapping colloid/fluid pairs on neighbour list
- calculate all other forces
- update velocities by a half step based on new forces
- apply collision procedure for velocities of particle pairs on neighbour list
- do physical measurements

Over a run of 500,000 steps of $\delta t = 0.01$, with 500 colloidal particles and 638 fluid particles in a box of size $20 \times 7 \times 7$, using *only* radial noise and friction, and coupling particle spin only via the tangential collision rule, we obtain a mean translational (kinetic) temperature $kT = 1.0012$. Over the ensemble of colloidal particles we obtain a spin temperature $T_s = 1.0006$. FIG. 4 shows the pair correlations compared to the Percus-Yevick theory of hard sphere mixtures.[41] Because the colloidal particles are slightly elastic, the correlation functions were calculated at a lower volume fraction $\phi^* = (r^*/R)^3 \, \phi = 0.2359$, where $r^*$ is the first solution of $g_{cc}(r^*) = 1$. For the colloid-colloid pair correlation $g_{cc}(r)$ the radial distance was rescaled by a factor $r^*/R = 0.9643$, but not for the other correlations because the particle-fluid interaction radius is $R = 0.5$. FIG. 4 shows an excellent match, the slight differences can be attributed to the softness of the particle-particle interaction. Hence, the temperature and distributions generated are correct within the statistical noise and the accuracy of the theory.

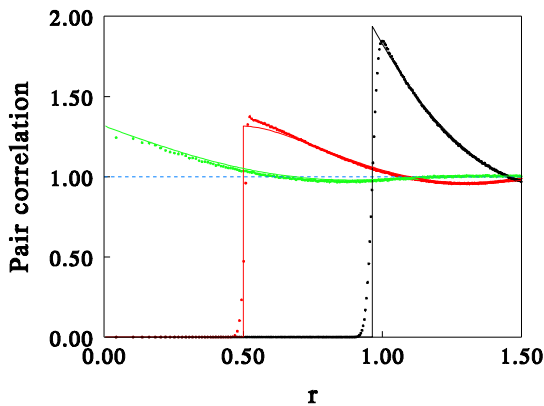

**FIG. 4** Pair correlations for $f = \mu = 1$, using collision rule Eq. (30) (dots) compared to rescaled Percus-Yevick theory for the pair correlation of a hard sphere mixture (lines). The three curves give the colloid-colloid (black), colloid-solvent (red) and solvent-solvent pair correlation (red).

Finally, we briefly study dynamics. How quickly is translational kinetic energy transferred to spin kinetic energy? This question is important as it implies how fast the spin of a particle can follow changes in the fluid rotation. To study this, we used again 500 colloidal particles and 638 fluid particles with *only* radial noise and friction ($f = 1$, $\mu = 0$). This is ordinary DPD, with an extra collision rule to couple spin and velocity. At time $t = 0$ all spins and velocities were reset to 0, and the increase of kinetic and rotational temperature were followed in time. This showed for the kinetic temperature an exponential increase in time with a relaxation time $\tau_k = 0.47 \pm 0.01$. If all moments of inertia are put to 1 we find a very slow increase of rotational temperature (or rotational correlation time), $\tau_r = 5.03 \pm 0.01$, i.e. an order of magnitude slower than the kinetic temperature relaxation. On dimensional grounds we can put $I = \lambda m R^2$, where $\lambda$ is a shape factor. If we vary $\lambda$ from 0.0625 to 4 we find a linear relation $\tau_r \approx 0.96(2)\lambda + 1.13(3)$. This means that even if the moment of inertia vanishes, the rotation correlation time is still twice the kinetic temperature relaxation time. Thus, the limiting factor for spin relaxation for small moments of inertia must be the collision frequency. As a compromise between fast rotational relaxation and reliable simulation (spin updates may become unstable for small moments of inertia) we choose $\lambda = 2/5$, the physical value for homogeneous solid bodies.

# V. APPLICATION TO COLLOID AND FIBER RHEOLOGY

To test the stick boundary condition and the modified friction force methods, systems comprising $N_c = 150$, 300, 500, 630, 750, 870, 100 or 1050 colloidal particles and $N_s$ solvent particles were simulated in a box of size $V = 20 \times 7 \times 7$. The solvent density between the colloid particles was fixed at $\rho r_c^3 = 3$ (i.e. $\rho = 0.8888$), hence the number of solvent particles was taken as $N_s = \rho V(1-\phi)$. Both colloid particles and solvent particles have diameter $D = 1$. The first set of runs test the application of the FPM method with modified shear friction. Using $f = \mu = kT = 1$ we obtained viscosity $\eta = 0.40$ for the pure solvent, hence using Eq. (26) we take $\mu^{CS} = 0.68$. Note that this viscosity is 20% *lower* than for $f = 1$, $\mu = 0$. This is probably due to fast diffusion, which increases viscosity at low friction.

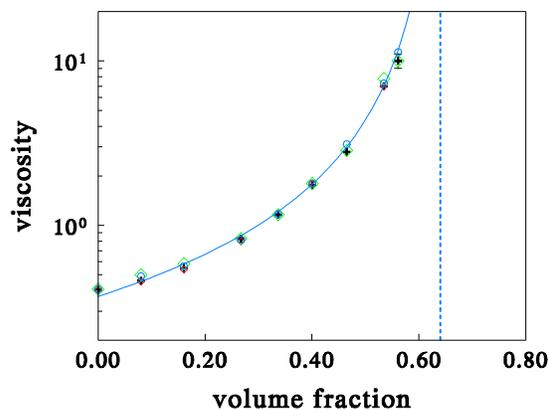

**FIG. 5** Viscosity of spherical particle suspensions for the modified friction method with Lowe-Anderson thermostat for step size $\delta t = 0.01$ (black crosses) and $\delta t = 0.005$ (red diamonds), radial reflection method (green diamonds) and for the tangential stick boundary condition method (blue circles).

FIG. 5 shows the simulation results for gravity field $g = 0.01$, which leads to a peak velocity $v_{max} = 0.27$ for the system without colloidal particles. The black dots show the simulation results for time step $\delta t = 0.01$ using the Lowe-Anderson thermostat for the radial interaction between colloid and solvent. The red dots give the results for time step $\delta t = 0.005$ and the green diamonds show the results for radial Lowe-Anderson replaced by radial reflection. In all cases the total simulation time was $t = 5000$. Finally, the blue circles show the results for the radial and tangential stick boundary condition method, again using step size $\delta t = 0.01$. All runs superimpose very well. The blue line is a fit to the Krieger-Dougherty equation, which for the stick boundary condition method leads to $[\eta] = 2.53 \pm 0.15$ and for the modified friction method gives $[\eta] = 2.49 \pm 0.13$. In both cases a maximum volume fraction $\phi_{max} = 0.640 \pm 0.015$ is obtained. These results show that both methods give reasonable results. There is no statistical difference between small or large steps, and there is no statistical difference between the two simulation methods.

To test the robustness of the method viscosity was increased in two ways. Firstly, the number density of the solvent particles was kept fixed at $\rho r_c^3 = 3$ (counting only the volume not occupied by colloidal particles), but the radial and shear friction factors were increased to $f = \mu = 3$. The time step taken was $\delta t = 0.01$, and systems were evolved over $3 \cdot 10^5$ steps. The black dots in FIG. 6a show the high viscosity results, compared to the lower viscosity results of FIG. 5. All data points have moved up in the graph by roughly the same amount, indicating the right scaling with solvent viscosity.

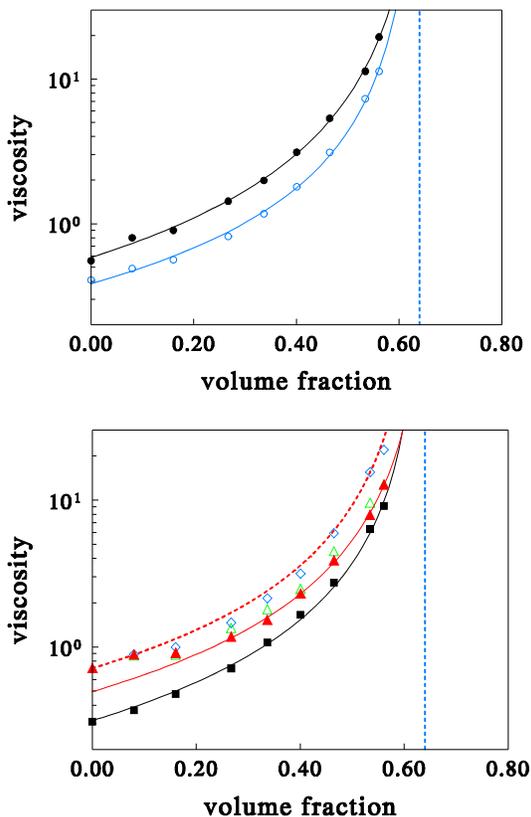

**FIG. 6a** (top) Viscosity of spherical particle suspensions for stick boundary condition method at $f = \mu = 1$ and $\rho r_c^3 = 3$ (blue circles) compared to $f = \mu = 3$ and $\rho r_c^3 = 3$ (black dots) **b** (bottom) Solvent density variation at $f = \mu = 1$ for $\rho r_c^3 = 2.2$ (black squares) and $\rho r_c^3 = 4.8$ (red triangles). Green triangles and blue diamonds show data for $\rho r_c^3 = 4.8$, but for $f^{CC} = \mu^{CC} = 2$ and 5 respectively.

When viscosity is increased by increasing solvent density, we see a different picture. FIG. 6b shows the results for $f = \mu = 1$, with solvent density $\rho r_c^3 = 4.8$ in red, as compared to $\rho r_c^3 = 2.2$ (in black). At low colloid volume fraction we find an increase in viscosity that is consistent with $[\eta] = 2.5$. The dashed red curve gives the fit to Eq. (2) with $[\eta] = 2.5$ and $\phi_{max} = 0.64$. Above volume fraction $\phi = 0.1$ the simulations cross over to the full red curve, which is the same curve but shifted down. The simulation results at high volume fraction seem to converge towards the full black curve that fits the high colloid fraction results with $\rho r_c^3 = 2.2$. This indicates that at high volume fraction, viscosity is dominated by direct colloid-colloid interactions, i.e. elastic repulsion and lubrication forces.

So far we have used the same radial and shear friction factors for solvent-solvent, colloid-solvent and colloid-colloid interaction. The result in FIG. 6b shows that the colloid-colloid friction must be treated with care to represent the lubrication forces correctly. The green triangles and blue diamonds in FIG. 6b show simulation results for $f^{CC} = \mu^{CC} = 2$ and 5 respectively. This shows that (at high volume fractions) we can roughly compensate for the loss in colloid-solvent interactions using $f^{CC} = \mu^{CC} \approx 5$. Near $\phi = 0.2$ viscosity is systematically below expectation, however, even if the colloid-colloid viscosity is increased. This may be related to the range of the direct friction interaction $r_c = 1.5$, which will be too small for $\phi < \phi_{max}/r_c^3 = 0.19$.

One may argue that friction between colloid and solvent (denoted by $f^{CS}$ and $\mu^{CS}$) should be set to $f^{CS} = \mu^{CS} = 0$, because this describes an "inert boundary". The choice so far has been $f^{CS} = f^{SS} = \mu^{SS}$, which describes a "thermostated boundary". Which is the preferred choice? Since torque and friction forces on a particle in an external flow field are clearly influenced by the colloid-solvent friction, one may wonder if colloid-solvent interaction is double counted if we use stick boundary conditions *and* use $f^{CS} = \mu^{CS} = f^{SS} = \mu^{SS}$. The low volume fraction results in FIG. 6 show that it doesn't matter. The physical significance of the boundary conditions is to fix the intrinsic viscosity $[\eta]$. Even though at higher volume fraction viscosity may cross-over to a lower lying Krieger-Dougherty curve, the result at low volume fraction $[\eta] = 2.5$ is quite robust.

The crossover to a lower lying viscosity curve at high volume fraction implies that high volume fraction systems have their own viscous behaviour that extrapolates to a different solvent viscosity $\eta_0$. In FIG. 6b this is illustrated by the full red curve (the high volume fraction result) that lies below the dashed red curve (the low volume fraction result). The difference between the two extrapolated values appears to increase with solvent density. So can we make the system self-consistent by choosing the right solvent density? From the Groot and Warren estimate of DPD viscosity[42] and the result by Marsh *et al*[31] for the kinetic part ($\eta_K = \rho D/2$), we expect $\eta \sim a + b(\rho r_c^3)^2$. Both the high and the low volume fraction results for $\eta_0$ follow this relation quite well. Solving for the solvent density $\rho^*$ where both viscosities are equal, we obtain the self-consistent point as

$$\begin{cases} \rho^* r_c^3 \approx 2.2 & if \quad f^{CS} = f^{CC} = f^{SS} \\ \rho^* r_c^3 \approx 2.0 & if \quad f^{CS} = 0; \, f^{CC} = f^{SS} \end{cases} \tag{41}$$

For the remainder of this study we will maintain as default $f^{CS} = f^{CC} = f^{SS}$, with the remark that there is room for optimisation of the colloid-colloid friction factor and range. Note again that $r_c = 1.5$ is the cut-off length of the friction interaction, hence the actual values to be used for $\rho^*$ in the present units (particle diameter = 1) are $\rho^* \approx$

0.65 for $f^{CS} = f^{CC} = f^{SS}$, and $\rho^* \approx 0.59$ for $f^{CS} = 0$ and $f^{CC} = f^{SS}$.

To test the simulation method for more complex cases, we now study the rheology of semi-flexible fibers. Systems were prepared with the same number of colloid particles as above in a volume V = 20×7×7. The colloid particles were bound together in strings of 2, 5 and 10 particles, thus forming semi-flexible fibers. The bending stiffness was taken as $K_b = 20$ (see Eq. (12)) and $E_{cc} = 10^4$. The particles were strung together by linear springs of spring constant C = 20-25. The value was chosen to obtain a mean centre-to-centre distance 1.

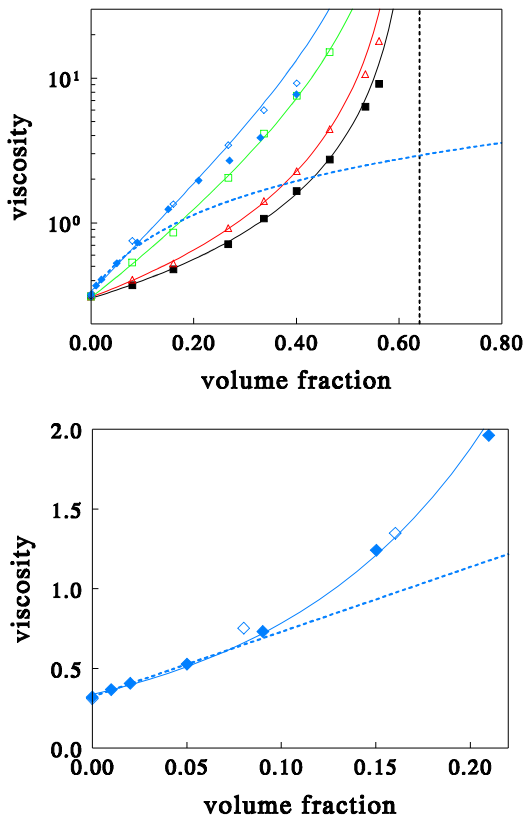

**FIG. 7a** (top) Viscosity of fiber suspensions for stick boundary condition method at $f = \mu = 1$ and solvent density $\rho = 0.65(1-\phi)$ for monomer particles (closed black squares), dimers (red triangles), 5-mers (open green squares) and 10-mers (blue diamonds). **b** (bottom) Enlargement of 10-mer data to linear scale for small system (open symbols) and large system (closed symbols).

The simulation results are shown in FIG. 7. The closed black squares show the data for monomers, as in FIG. 6b. The red, green and blue symbols and curves show the results for dimers, 5-mers and 10-mers respectively. The dimer and 5-mer data are fitted well by the Krieger-Dougherty equation, Eq. (2). The intrinsic viscosities are determined from the first two points ($\phi = 0$ and $\phi = 8\%$). For a linear fit we obtain for dimers: $[\eta] = 3.8$ and for 5-mers: $[\eta] = 9.0$. If we fit the first two points to the Krieger-Dougherty equation, assuming $\phi_{max} = 0.64$, we find $[\eta] = 3.1$ for dimers, and $[\eta] = 6.3$ for 5-mers respectively. This gives the simulation estimates $[\eta] = 3.5\pm0.3$ for dimers, and $[\eta] = 7.7\pm1.3$ for 5-mers. For touching dimers the exact result is $[\eta] = 3.45$ and for ellipsoids of $A = 5$, $[\eta] = 8.1$. These values compare well with the simulation. This result also shows that for long fibers that we need low volume fractions for an accurate determination of the intrinsic viscosity.

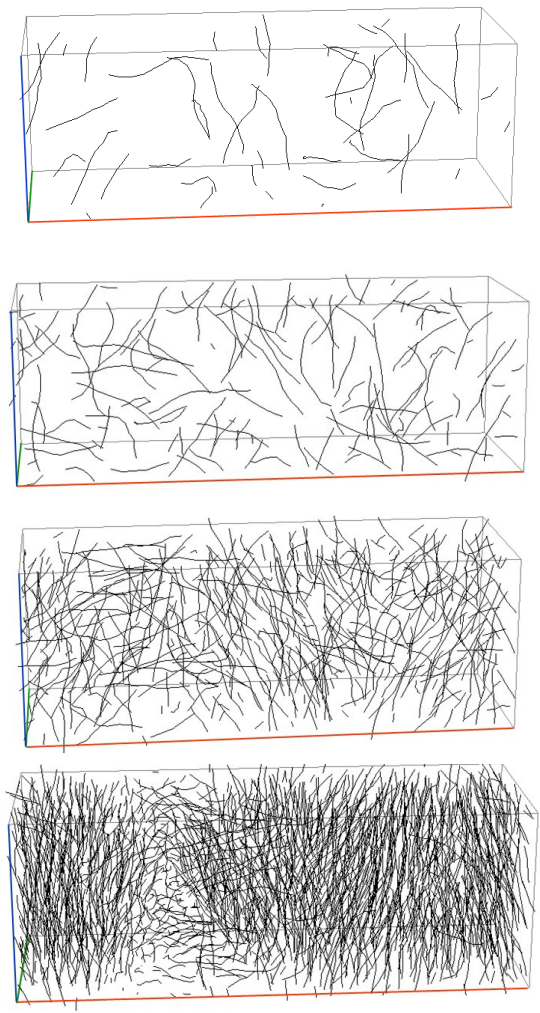

**FIG. 8** Conformations of sheared semi-flexible 10-mers, top to bottom: $\phi = 0.02$, 0.05, 0.15 and 0.40.

The 10-mer data tends to curve down to the 5-mer data near volume fraction $\phi = 0.4$. For this length the chains show some flexibility; the typical end-to-end distance is 8.3 (rather than 9), so this means that the fiber size is comparable to the system size. To test whether the observed behaviour is a finite size artefact the system size was doubled in all dimensions to V = 40×14×14, containing 5000 to 9000 particles. This also allows exploration of lower volume fraction, down to 1% (30 10-mers and 5008 solvent particles). A gravity field $g = 0.01$ (see Eq. (21)) was applied to the colloid particles, and $g = 0.0034$ was applied to the solvent particles. The results are shown in FIG. 7a by the full blue diamonds, and an enlargement to linear scale is shown in FIG. 7b. The linear curve is a fit to the first four points, giving an intrinsic viscosity [η] = 12.5±0.4. The theoretical value obtained from Eq. (3), for an ellipsoid of aspect ratio $A$ = 9.3 is 12.3, an excellent match with the simulation result.

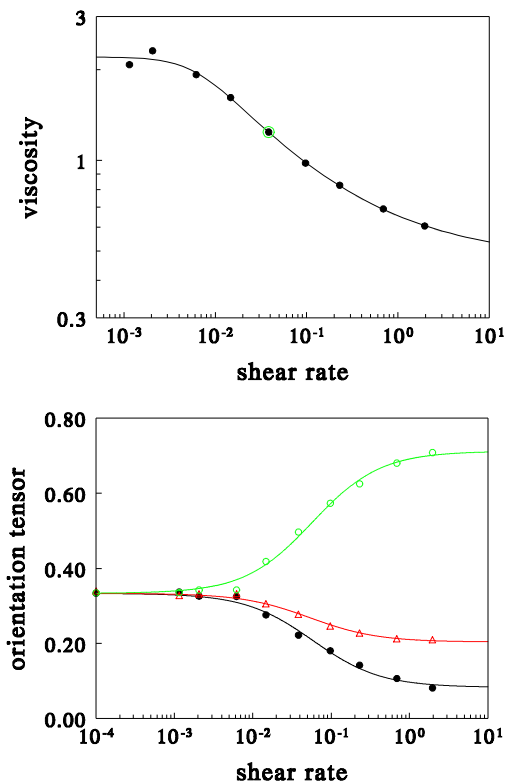

**FIG. 9a** (top) Viscosity in Poisseuille flow for 15% volume fraction semi-flexible fibers of 10 beads. The point corresponding to the driving force applied in FIG. 7 is marked by the green circle. The black line is a fit to the Carreau model. **b** (bottom) Orientation tensor obtained from end point vectors, normalised to 1, for $xx$-component (black dots), $yy$ (red triangles) and $zz$ (green circles).

To investigate if the relatively low viscosity for 10-mers shown in FIG. 7a may be caused by nematic ordering and consequently shear thinning, the conformations of four systems are shown in FIG. 8. The systems chosen have particle volume fractions φ = 0.02, 0.05, 0.15 and 0.40. Note that the left half of the systems is pushed up, while the right half is being pushed down. Hence the areas of high shear are the middle of the box and the left and right edges. This clearly shows that no nematic order is visible at 2% volume fraction; some order can be seen at 5%, while the system at 40% is fully ordered. We find exactly parabolic velocity profiles for the highest and the lowest volume fraction, but some deviation from a parabolic flow profile for φ = 0.15. This suggests that this system is most shear-thinning of the samples shown.

To test this hypothesis, a random conformation was generated in the isotropic state for φ = 0.15, and this conformation was copied eight times. Then each copy was evolved over $10^5$ time steps (δ$t$ = 0.01) to attain a stationary state, and the velocity profile was measured over the next $4 \cdot 10^5$ time steps. The driving force on the colloid particles was varied as $g_C$ = 0.0005, 0.001, 0.0025, 0.005, 0.01, 0.02, 0.04, 0.1 and 0.25. In all cases the driving force on the solvent was chosen as $g_S$ = 0.34 $g_C$. This indeed shows that the sample at φ = 0.15 and driving force $g_C$ = 0.01 is shear-thinning. The *maximum* shear rate in the system was calculated from the velocity profile; viscosity is plotted against this shear rate in FIG. 9. The system corresponding to the shear rate of FIG. 7 is marked by a green circle. Near this point we obtain a slope $\eta \propto \dot{\gamma}^{-0.27\pm0.1}$. A more comprehensive analysis, taking all points into account by fitting to $\eta = (\eta_0 - \eta_\infty)/(1 + (\tau\dot{\gamma})^2)^{(1-n)/2} + \eta_\infty$ (the Carreau model) gives the black line in FIG. 9a, with Newtonian index $n$ = 0.56±0.08 and relaxation time τ = 170±50. Note that the first point is quite uncertain, if this is left out of the statistics we find $n$ = 0.60±0.03 and τ = 250±40.

To further characterise the structure of the system the endpoint separation vector $\mathbf{r}_e = (x_e, y_e, z_e)$ was sampled for each polymer, and the ensemble average was taken for the $xx$, $yy$ and $zz$ components of the orientation tensor $T_{xx} = \langle x_e^2 \rangle / \langle \mathbf{r}_e^2 \rangle$, $T_{yy} = \langle y_e^2 \rangle / \langle \mathbf{r}_e^2 \rangle$, $T_{zz} = \langle z_e^2 \rangle / \langle \mathbf{r}_e^2 \rangle$. These results are shown in FIG. 9b. This shows that the system is nearly fully ordered by $\dot{\gamma} \approx 2$, or $\tau\dot{\gamma} \approx 300 - 500$. It is interesting to notice that in the $x$-direction the polymers order over a narrower range than in the $y$-direction, as $T_{xx}<T_{yy}$. Note that $x$ is the direction of the shear gradient, whereas there in no gradient in the $y$-direction.

# VI. SUMMARY AND CONCLUSIONS

Some important points of the theory of colloid rheology are reviewed, and based on literature results an accurate expression is given for the intrinsic viscosity of suspensions of ellipsoids, covering the whole shape range from flat discs to sharp needles. The Fluid Particle Model (FPM) is briefly reviewed, which comprises radial and transverse noise and friction between simulated particles; and a method is given to simulate semi-flexible fibers.

The simplest approach to model long-range hydrodynamics is to use one particle type as solvent for colloid particles, and to include *only* viscous interaction between solvent and colloid. This is the free draining approximation. The intrinsic viscosity obtained this way is not reliable, however. It results from the increased density of friction centres when particles are added to a heat bath, and not from the physical cause – the fluid boundary conditions at the particle surface.

This conclusion forces the study how to impose the fluid boundary conditions. Two methods are considered. In the first method radial and transverse friction forces between particle and solvent are applied such that the correct friction and torque follows for moving or rotating particles. The force coefficients are calculated analytically and checked by numerical simulation. In the second method collision rules are derived between colloidal particle and solvent particle that impose the stick boundary conditions exactly. The collision rule comprises a generalisation of the Lowe-Anderson thermostat for transverse velocities.

When applied to colloid rheology it was found that, although the intrinsic viscosity is correct by construction, viscosity at high volume fraction can be lower than expected from the low volume fraction results. The difference varies with solvent density. The low and high volume fraction viscosities can be made self-consistent by choosing the correct solvent density. It is expected that this self-consistent solvent density depends on the range of the colloid-colloid friction interaction, and it is shown that it depends on the colloid-solvent friction.

When applied to semi-flexible fibers of various lengths, the simulations give intrinsic viscosities in line with the theoretical values. The results are well fitted by the Krieger-Dougherty equation, but some deviations are found for fibers of 10 beads. These are caused by shear thinning, which in turn is related to an isotropic-nematic transition under shear.

This opens the way to investigating the rheology of fiber mixtures with full hydrodynamic interactions. With the method described we can study how viscosity and shear-thinning behaviour depend on fiber length distribution and on the persistence length of the fibers. The method is suitable to deal with high volume fractions and entanglements, and since it is based on explicit simulation associative interactions between fibers can be included. Finally, the transverse collision rule given here may be generalised to collisions where both particles carry spin. The collision rule is also valid for non-ideal solvents, but in that case care must be taken to prevent (artificial) solvent-induced colloidal interactions.

## APPENDIX A: DERIVATION OF RADIAL AND SHEAR FRICTION COEFFICIENTS

We start with the shear friction coefficient. The flow field around a sphere of radius $R$ and with spin $(0, 0, \omega)$ is given by[33]

$$\mathbf{v} = \frac{\omega R^3}{r^2} \sin \theta \, \mathbf{e}_\phi \qquad (A1)$$

The outer product in Eq. (25) is worked out as $\mathbf{r}_{ij} \times \boldsymbol{\omega}_{ij} = -\omega r \sin \theta \, \mathbf{e}_\phi$, hence the shear force on the rotating sphere due to a neighbouring fluid particle follows as

$$\mathbf{F}^S = -\mu^{CS} R(p - r/R)^2 \theta(p - r/R) \\ \cdot \omega R \sin \theta \left(R^2/r^2 - r/R\right) \mathbf{e}_\phi \qquad (A2)$$

The torque on the particle (see Eq. (9)) is thus obtained as

$$\tau_z = -\int (\mathbf{r}_{ij} \times \mathbf{F}^S)_z \rho d^3 \mathbf{r}_{ij}$$

$$= -\int r \sin \theta \left| F^S \right| \rho \, r^2 dr \sin \theta \, d\theta \, d\phi$$

$$= -2\pi \mu^{CS} \rho \omega R^6 \int_1^p x^3 (p - x)^2 (1/x^2 - x) dx \qquad (A3)$$

$$\cdot \int_{-1}^{1} \sin^2 \theta \, d\cos\theta$$

$$= -\left(\frac{1}{7} - \frac{4}{9} p + \frac{2}{5} p^2 - \frac{1}{9} p^4 + \frac{4}{315} p^7\right) \cdot 2\pi \mu^{CS} \rho \omega R^6$$

Equating the calculated torque to the theoretical torque $\boldsymbol{\tau} = -8\pi \eta R^3 \boldsymbol{\omega}$ for stick boundary conditions, we find the shear friction factor as

$$\mu^{CS} = \frac{1260}{45 - 140p + 126p^2 - 35p^4 + 4p^7} \cdot \frac{\eta}{\rho R^3}$$

$$= \frac{105}{556} \cdot \frac{\eta}{\rho R^3} \approx 0.18885 \frac{\eta}{\rho R^3} \qquad (A4)$$

where $\rho$ is the density of solvent particles and $\eta$ is the solvent viscosity, and where $p$ is the cut-off range in Eq. (25). The numerical value is given for range $p = 3$ that is used here.

The force on a moving particle can be calculated in a similar way using the known solution of the flow field.[33] We start with a particle moving in the positive $z$-direction with velocity $u$ and then transform to a co-moving frame where the particle is at rest. In this frame we have the flow field

$$\mathbf{v} = u \cos\theta \left[ -1 + \frac{3}{2} \frac{R}{r} - \frac{1}{2} \left(\frac{R}{r}\right)^3 \right] \mathbf{e}_r \\ - u \sin\theta \left[ -1 + \frac{3}{4} \frac{R}{r} + \frac{1}{4} \left(\frac{R}{r}\right)^3 \right] \mathbf{e}_\theta \qquad (A5)$$

Next, the force on the colloid particle is $\mathbf{F} = f^{CS} R W(r) v_r \mathbf{e}_r + \mu^{CS} R W(r) v_\theta \mathbf{e}_\theta$, hence the force component in the $z$-direction is $F_z = f^{CS} R W(r) v_r \cos\theta - \mu^{CS} R W(r) v_\theta \sin\theta$, where $W(r) = (p - r_{ij}/R_i)^2 \, \theta(p - r_{ij}/R_i)$ is the distance dependence of the friction force. Integrating over all neighbouring particles we find for a general range

$$F_z = 2\pi \rho u \int_{-1}^{1} d\cos\theta \int_R^{pR} r^2 R W(r)$$

$$\cdot \left\{ f^{CS} \cos^2 \theta \left[ -1 + \frac{3}{2} \frac{R}{r} - \frac{1}{2} \left(\frac{R}{r}\right)^3 \right] \right.$$

$$\left. + \mu^{CS} \sin^2 \theta \left[ -1 + \frac{3}{4} \frac{R}{r} + \frac{1}{4} \left(\frac{R}{r}\right)^3 \right] \right\} dr$$

$$= \frac{4\pi}{3} \rho u R^4 \int_1^p x^2 (p - x)^2$$

$$\cdot \left\{ f^{CS} \left[ -1 + \frac{3}{2x} - \frac{1}{2x^3} \right] + 2\mu^{CS} \left[ -1 + \frac{3}{4x} + \frac{1}{4x^3} \right] \right\} dx$$

$$= -\frac{4\pi}{3} \rho u R^4$$

$$\cdot \left\{ f^C \left[ \frac{1}{2} p^2 \ln p + \frac{1}{30} p^5 - \frac{1}{8} p^4 - \frac{1}{3} p^2 + \frac{1}{2} p - \frac{3}{40} \right] \right.$$

$$\left. + \mu^{CS} \left[ -\frac{1}{2} p^2 \ln p + \frac{1}{15} p^5 - \frac{1}{8} p^4 + \frac{5}{6} p^2 - p + \frac{9}{40} \right] \right\}$$

$$(A6)$$

Equating this to the known friction force $\mathbf{F} = -6\pi \eta R \mathbf{v}$, and substitution of the previous result for the shear friction coefficient, Eq. (A4), we find for a force range $p = 3$

$$f^{CS} = \frac{4725 \cdot \ln 3 + 13680}{25020 \cdot \ln 3 - 20016} \cdot \frac{\eta}{\rho R^3} \approx 2.52580 \frac{\eta}{\rho R^3} \quad (A7)$$